\documentclass[journal]{IEEEtran}
\ifCLASSINFOpdf
\else
\fi

\usepackage{amsmath}
\usepackage{rotating}
\usepackage{subcaption}

\usepackage{array}
\usepackage[linesnumbered,ruled,vlined]{algorithm2e} 
\usepackage{amsmath}      
\usepackage{amssymb}      
\usepackage{mathtools}    
\usepackage{booktabs}   
\usepackage{todonotes}
\usepackage[normalem]{ulem}
\usepackage{hyperref}

\usepackage[table]{xcolor}
\newcommand{\good}[1]{\textcolor{green!60!black}{#1}}
\newcommand{\bad}[1]{\textcolor{red!70!black}{#1}}

\begin{document}
%
\title{Simopt-Power: Leveraging Simulation Metadata for Low-Power Design Synthesis}
%
%
%
\author{\IEEEauthorblockN{Eashan Wadhwa \& Shanker Shreejith}\\
\IEEEauthorblockA{Department of Electronic and Electrical Engineering,
Trinity College Dublin\\
Dublin, Ireland\\
\{wadhwae, shreejith.shanker\}@tcd.ie}}
\maketitle

\begin{abstract}
Excessive switching activity is a primary contributor to dynamic power dissipation in modern FPGAs, where fine-grained configurability amplifies signal toggling and associated capacitance. Conventional low-power techniques -- gating, clock-domain partitioning, and placement-aware netlist rewrites -- either require intrusive design changes or offer diminishing returns as device densities grow. In this work, we present Simopt-power, a simulator-driven optimisation framework that leverages simulation analysis to identify and selectively reconfigure high-toggle paths. By feeding activity profiles back into a lightweight transformation pass, Simopt-power judiciously inserts duplicate truth table logic using Shannon Decomposition principle and relocates critical nets, thereby attenuating unnecessary transitions without perturbing functional behaviour. We evaluated this framework on open-source RTLLM benchmark, with Simopt-power achieves an average switching-induced power reduction of \(\approx\)9\% while incurring only \(\approx\)19\% additional LUT-equivalent resources for arithmetic designs. These results demonstrate that coupling simulation insights with targeted optimisations can yield a reduced dynamic power, offering a practical path toward using simulation metadata in the FPGA-CAD flow.
\end{abstract}

%
\IEEEpeerreviewmaketitle

\section{Introduction}
Field-programmable gate arrays (FPGAs) have emerged as critical components in modern digital systems, offering flexible hardware reconfigurability, rapid prototyping capabilities, and optimised parallel execution. As these devices increasingly permeate various application domains -- from embedded systems to high-performance computing -- power efficiency and reduction in switching activity have become paramount design considerations. Switching activity, closely correlated with dynamic power dissipation, significantly impacts FPGA reliability, performance, and overall system longevity. FPGA design tools play a pivotal role in managing and minimising this switching activity through optimisation algorithms and resource allocation strategies during synthesis, mapping, placement, and routing phases. Dynamic power in FPGAs arises predominantly from switching activity, with factors such as logic transitions, clock distribution networks, and signal routing contributing significantly.

Despite the growing sophistication of FPGA tools, effectively managing switching activity remains a substantial challenge, intricately tied to the increasing complexity of devices, shrinking technology nodes, and demanding performance constraints. For decades, efforts to curb dynamic power in FPGAs have centred on fine-grained techniques such as dynamic voltage and frequency scaling (DVFS) \cite{madala2024energy}, power gating of idle regions \cite{jahanirad2023dynamic}, and clock gating based on toggling statistics—strategies \cite{attaoui2021clock} now embedded in most modern FPGA-CAD tools. Yet these approaches focus exclusively on controlling activity \textit{after} a design is committed to hardware; they do not exploit insights available during simulation.  In every hardware designer's workflow, functional simulation serves solely to validate correctness and is subsequently discarded, leaving a rich trove of toggle-rate information untapped for power optimisation.

This paper addresses this gap by analysing the efficacy of modern FPGA synthesis and placement-and-routing tools in minimising switching activity through simulations, thereby optimising dynamic power usage. The specific contributions of this work are outlined below: 
\begin{itemize}
\item We extend an existing simulation infrastructure to harvest fine-grained switching-activity metadata, yielding per-signal toggle estimates across the entire design.
\item We feed these activity profiles into a new truth-table decomposition engine that explicitly balances power reduction against the accompanying area overhead.
\item We validate the approach on cutting-edge, large-language-model-generated Verilog suites (RTTLM \cite{lu2024rtllm}) alongside the standard VTR benchmarks \cite{Koios}, confirming significant power savings with acceptable area trade-offs.
\end{itemize}
We present a modest switching-induced power reduction of \(\approx\)9\% while incurring only \(\approx\)19\% additional LUT-equivalent resources for arithmetic designs. While providing empirical insights into the impact of tool-driven optimisations, evaluate their trade-offs against performance metrics, and suggest future directions for FPGA tool development to achieve enhanced power-aware computing. 

\section{Preliminaries}
\subsection{Verilog-To-Routing and Simulators}
\begin{figure*}[h]
  \centering
  \includegraphics[width=\textwidth]{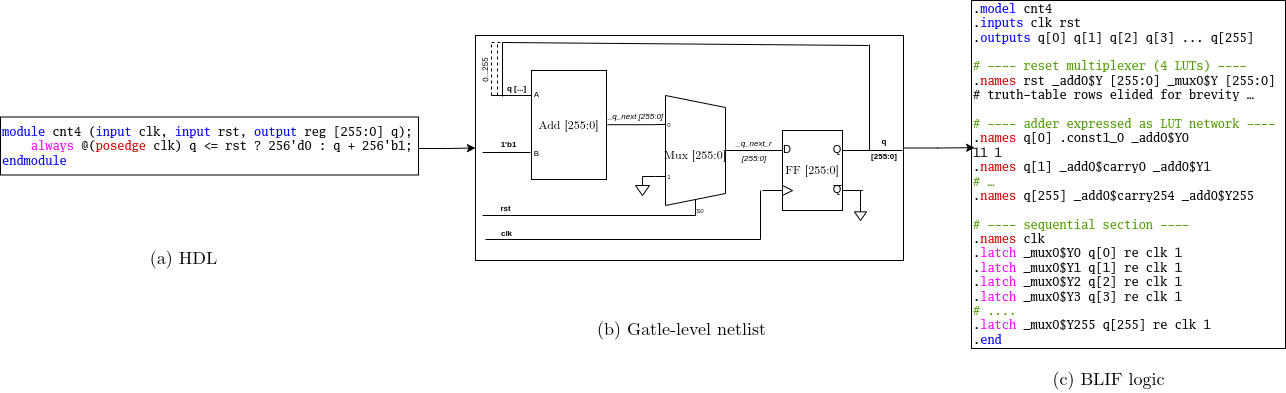}
  \caption{A standard Yosys synthesis flow, where HDL is synthesised to a gate-level netlist, followed by the generation of a BLIF logic netlist.}
  \label{fig:synthesisflow}
\end{figure*}
The open-source Verilog-To-Routing \cite{elgammal2024vtr} toolflow has been a key enabler for research into CAD optimisations for FPGAs and ASICs. 
It benefits from open-source community backing and seamless integration into custom and vendor toolflows, while delivering competitive power, area, and performance results for FPGA/ASIC designs \cite{shi2024open, li2024dag, wadhwa2024simopt}. Many researchers have expanded the original VTR flow with custom extensions for novel architecture exploration, enhanced routing algorithms and others.
In~\cite{shi2024open}, the authors enhanced the existing VTR flow for supporting a custom architecture which replaces expensive logic (connection, switching) blocks. The authors in~\cite{li2024dag} implemented a fine-grained optimisation inside Berkley's ABC \cite{barzen2023narrowing} on Intermediate representations (IR) graphs to lower the area incurred by designs mapped through VTR. 
The VTR backend is organised as a modular collection of toolsets, each implementing a distinct stage of the FPGA CAD flow—from logic synthesis through packing, placement, routing, and timing analysis—thereby enabling an end-to-end, open-source evaluation framework \cite{elgammal2024vtr}. This openness has empowered researchers to prototype and evaluate cutting-edge optimisation algorithms that drive measurable gains across critical performance metrics. A recent example is MapTune \cite{liu2025maptune}, which proposes a reinforcement learning framework for guided mapping of cells during the synthesis step, reducing delay and area by a considerable amount. Simopt \cite{wadhwa2024simopt} introduces simulation metadata into the FPGA-CAD synthesis flow, emitting lower latency designs for FPGAs. Using simulations in accelerating efficiency has also been studied in other domains. Combinational equivalence checking  of circuit networks is accelerated by studying simulation patterns \cite{rizzi2025simgen} markedly reducing SAT-solver runtime and verification effort. FPGA-Tidbits \cite{jellum2023fpga}, a Chisel-based library generates fully synthesisable RTL and employs Verilator for cycle-accurate co-simulation, enabling rapid co-design and prototyping of area-efficient FPGA accelerators .
Previous research in simulators however have never addressed the problem of reducing power consumption in generated designs, which our work addresses.

\subsection{FPGA netlist to placed design}
Fig.~\ref{fig:synthesisflow} shows canonical front-end flow of how synthesis tools input a Hardware-Description Language (HDL) snippet to emitting a nelist. After parsing the HDL, a boolean network is created which map arithmetic operators, muxes and comparators to logic operations while having no knowledge of platform specific cell libraries. This can then be directly mapped to gate-level netlists, shown as \textit{(b)} in Fig.~\ref{fig:synthesisflow} while incorporating gate-sizing, and library patterns for k-input Look-up tables (LUTs). The final step of netlist emission, \textit{(c)} in Fig.~\ref{fig:synthesisflow}, includes cell libraries to map the gate-level netlist to capture the technology-mapped boolean network.  In our work we use the standard open-source front-end for FPGA research flows - Yosys, which parses into an internal RTL Intermediate Language (RTLIL) representation, an equivalent of a boolean network. A series of optimisations are performed on this internally represented abstract tree, followed by mapping which replaces the RTLIL primitives with ABC's gate library to generate out the gate-level netlist. It is then incrementally lowered to a generic BLIF netlist through structural hashing and cut enumeration implemented in Berkeley's ABC. This netlist containing LUT, FF and latch directives can then be fed into placement tools for \textit{placing} the design onto an FPGA platform.
Placement determines the physical coordinates on the FPGA fabric contained in the design netlist. Placement typically is directed by timing closure, dynamic power and congestion factor. Modern placers integrate routability estimators that penalize hot-spots of overlapping demand, while power-aware heuristics redistribute high dynamic power consuming nets to shorten high-activity wires and avoid thermal concentration. However by this stage, there is very little a designer can do to reduce switching activity for designs. Simopt-power mitigates this issue by judiciously reducing the capacitance associated with high-toggling signals in-turn reducing the dynamic power consumption as we explain later in this work.
But before the resource mapping is solidified through synthesis and the spatial arrangement finalised by placement, we first turn to simulations to verify that a configured design meets all functional and timing requirements before committing it to hardware.
\subsection{Simulation in Simopt}
Simulations offer a virtual environment where designs are exercised under stimuli, enabling early verification and correctness of functionality before any hardware resources are committed. Although designers can choose from a range of commercial and open-source simulators, these tools typically fail to expose the internal metrics necessary to inform subsequent synthesis and placement stages.

\begin{figure}[h]
  \centering
  \includegraphics[width=0.48\textwidth]{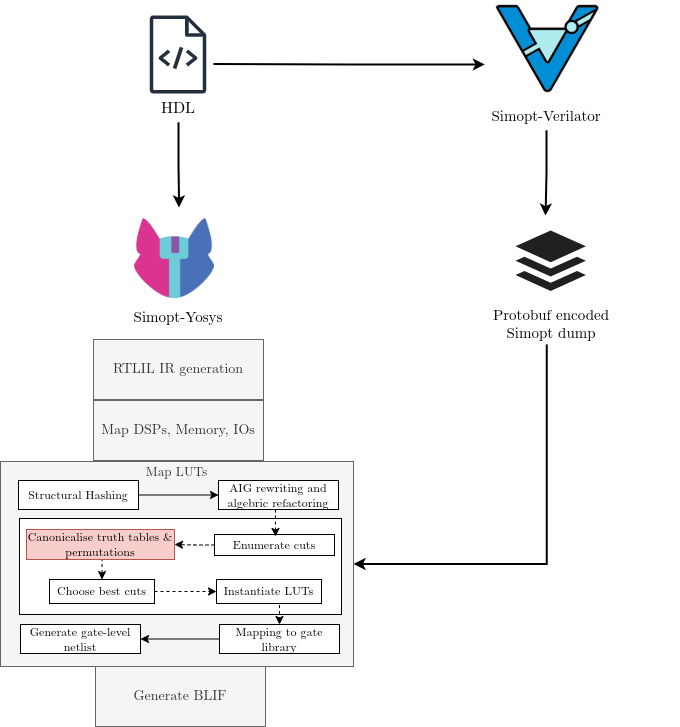}
  \caption{How Simopt-Power integrates into the Simopt framework. The red box indicates the step where the proposed truth-table decomposition step is integrated.}
  \label{fig:simopt-power-flow}
\end{figure}

In \cite{wadhwa2024simopt} however, authors introduced a Simopt framework which have modified Verilator to extract simulation metadata from a  design to generate \textit{simopt-counters} (shown in Fig.~\ref{fig:simopt-power-flow}, as the \textit{Protobuf-encoded Simopt dump}). \textit{Simopt-counters} tracks number of times a simulated net is toggled for a given set of input stimuli, which was later used by various \textit{Simopt-backends}. Authors demonstrated this framework by integrating these counters into one such open-source framework, Yosys to generate lower latency designs. However, their findings only focused on one such backend and only combinatorial circuits only. In our work, we re-use these simopt-counters as \textit{switching activity} estimates since they essentially give similar counts to toggling activity of a net.
We explain how we re-used this framework in the following section~\ref{sec:explanation}.
\section{Simopt-power}\label{sec:explanation}
Power consumption on modern FPGAs is primarily composed of static power and dynamic power, with the latter contributed by the dynamic switching activity of the active logic elements. 
Dynamic power on an FPGA can be captured through the switching power equation $$
\approx \alpha \cdot C \cdot V^2 \cdot f
$$
where $C$, $V$, $f$ and  $\alpha$ represent effective switch capacitance, supply voltage, clock frequency and switching factor (i.e. the average number of signal transitions per node of a circuit), respectively. Dynamic power consumption optimisation thus requires the tools to minimise the impact of each of these parameters where possible. $V$ is a device-level parameter which can have a significant impact on signal integrity, whereas in high-performance designs, $f$ should be as high as possible. Prior research has explored the possibility of dynamic voltage frequency scaling (DVFS)~\cite{Zhao2016} and clock gating to optimise power consumption at runtime using offline characterisation of designs~\cite{Zhao2016}. 
Our approach is to optimise the design-time parameters $\alpha$ and $C$ to minimise the power consumption on the logic when deployed on the FPGA. 
By duplicating the driving logic, each copy of the driving logic can be placed physically close to a subset of the loads, reducing the wire length to drive the logic to result in a smaller overall $C$, and/or reducing the number of programmable interconnect switches that the logic needs to pass through, resulting also in lower overall $C$. The duplication of driving logic, as we also observe from our results, typically results in an additional few \textmu W's, which is outweighed by the drop in dynamic power \cite{madala2024energy,jahanirad2023dynamic,attaoui2021clock}. 

To determine the signal toggle levels under actual input conditions, the design is initially simulated with Simopt-Verilator with a test fixture that replicates real-world data that the system could observe. 
Simopt-Verilator records per-cycle toggle counts, referred to as \textit{simopt-counters}, for every signal and register (entity).  These activity profiles are serialised as Protobuf-encoded metadata files and later parsed by Simopt-Power, which uses the counters to drive its power-reduction transformations.
The \textit{simopt-counters} tags each signal selected in the design, and its value is incremented each time the signal state toggles when the design is simulated using the Simopt-Verilator framework
To illustrate this flow, consider the simple example in Fig.~\ref{fig:synthesisflow}, where a free-running clock is fed as an input clock $clk$ for 512 ticks (Fig.~\ref{fig:synthesisflow}(a)). The gate-level netlist of the counter, shown in ~\ref{fig:synthesisflow}(b), uses a multi-bit vector $q$ to store the counter's state whose bit position toggles at the clock edge depending on the current state, and is individually tracked by the corresponding \textit{simopt-counter}. The internal next-state wires, $\_q\_next$, $\_q\_next\_r$ shown in Fig.~\ref{fig:synthesisflow}(b) are also tracked by separate \textit{simopt-counters}. During the synthesis flow, the relation between \textit{simopt-counters}, Verilator's intermediate signal representations, and Yosys's synthesised net names are maintained by the Simopt-Verilator flow. 


\begin{algorithm}
\caption{\textsc{TruthTableDecompose} -- Shannon-Decomposes a truth-table}
\label{alg:ttsplit}
\SetAlgoLined
\KwIn{%
  $\mathsf{T}$ -- pointer to the truth table on the \emph{right} cut
  of size $n_{\text{right}}$ variables\;
  $\mathcal{C}_{\text{left}}$ -- indices of the $n_{\text{left}}$ variables
  in the cut\;
  $\mathcal{C}_{\text{right}}$ -- indices of the $n_{\text{right}}$ variables in the cut\;}
\KwResult{$\mathsf{T}$ is replaced by the Shannon-decomposed truth-table with positive and negative co-factors.}

\BlankLine
\If{\textbf{not} $\mathcal{C}_{\text{simopt\_counter}} < \mathit{simopt\_counters\_median}$}{
  \Return
}

\For{$i \gets 0$ \KwTo $n_{\text{right}}-1$}{
  $\mathit{splitVar} \gets -1$\;
   \If{$\mathcal{C}_{\text{right}}[i] \notin \mathcal{C}_{\text{left}}$}{
       $\mathit{splitVar} \gets \mathcal{C}_{\text{right}}[i]$; break\;
   }
  \If{$\mathit{splitVar} = -1$}{
  \textbf{continue}\tcp*[r]{nothing to split}
  }

$pos \gets$ index of $\mathit{splitVar}$ in $\mathcal{C}_{\text{right}}$\;

$n_{\text{words}} \gets 2^{\,n_{\text{right}}}/(8\cdot\texttt{sizeof(word)})$\;
allocate arrays $\mathsf{T}_0,\mathsf{T}_1$ of $n_{\text{words}}$ words\;

$\mathsf{T}_0 \gets \textsc{ShannonCofactor}(\mathsf{T}, n_{\text{right}}, pos, 0)$\;
$\mathsf{T}_1 \gets \textsc{ShannonCofactor}(\mathsf{T}, n_{\text{right}}, pos, 1)$\;
copy ${\mathsf{T}_1, \mathsf{T}_0} $ into $\mathsf{T}$\;

free $\mathsf{T}_0,\mathsf{T}_1$\;
}
\For{$i \gets 0$ \KwTo $n_{\text{left}}-1$}{
$\mathit{splitVar} \gets -1$\;

   \If{$\mathcal{C}_{\text{left}}[i] \notin \mathcal{C}_{\text{right}}$}{
       $\mathit{splitVar} \gets \mathcal{C}_{\text{left}}[i]$; break\;
   }
  \If{$\mathit{splitVar} = -1$}{
  \textbf{continue}\tcp*[r]{nothing to split}
  }

$pos \gets$ index of $\mathit{splitVar}$ in $\mathcal{C}_{\text{left}}$\;

$n_{\text{words}} \gets 2^{\,n_{\text{left}}}/(8\cdot\texttt{sizeof(word)})$\;
allocate arrays $\mathsf{T}_0,\mathsf{T}_1$ of $n_{\text{words}}$ words\;

$\mathsf{T}_0 \gets \textsc{ShannonCofactor}(\mathsf{T}, n_{\text{left}}, pos, 0)$\;
$\mathsf{T}_1 \gets \textsc{ShannonCofactor}(\mathsf{T}, n_{\text{left}}, pos, 1)$\;
copy ${\mathsf{T}_1, \mathsf{T}_0} $ into $\mathsf{T}$\;

free $\mathsf{T}_0,\mathsf{T}_1$\;
}

\end{algorithm}
\begin{figure}[t]
  \centering
  \includegraphics[width=0.42\textwidth]{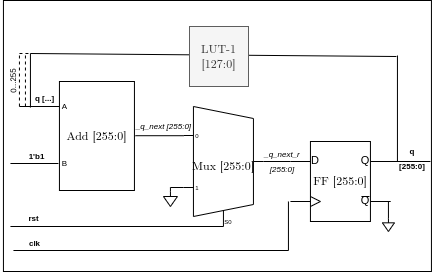}
  \caption{Simopt-power truth-table decomposed synthesis netlist of the counter from Fig.\ref{fig:synthesisflow}(b). The grey box indicates the additional LUT logic added by the Simopt-Power framework to reduce power.}
  \label{fig:synthesisflowdecompose}
\end{figure}

\begin{algorithm}
\caption{\textsc{ShannonCofactor} -- Calculates Shannon cofactor of truth-table}
\label{alg:cofactor}
\SetAlgoLined
\KwIn{
  $\mathsf{T}_{\text{src}}$ -- pointer to the original truth table on $n$ variables\;
  $n$ -- number of variables\;
  $p$ -- position of variable to fix\;
  $v$ -- value to cofactor on ($v \in \{0,1\}$)
}
\KwOut{
  $\mathsf{T}_{\text{dst}}$ -- cofactor of $\mathsf{T}_{\text{src}}$ with $x_p = v$
}

\BlankLine
$n_{\text{words}} \gets 2^n / (8 \cdot \texttt{sizeof(word)})$\;
\For{$i \gets 0$ \KwTo $n_{\text{words}}-1$}{
    $\mathsf{T}_{\text{dst}}[i] \gets 0$\;
}

\tcp*[l]{Scan every minterm and check if it matches $x_p = v$}
\For{$m \gets 0$ \KwTo $2^n - 1$}{
    \If{bit $m$ is set in $\mathsf{T}_{\text{src}}$}{
        \If{$(m \gg p)\ \&\ 1 = v$}{
            $m' \gets$ remove bit $p$ from $m$ (collapse to $n-1$ vars)\;
            set bit $m'$ in $\mathsf{T}_{\text{dst}}$\;
        }
    }
}
\end{algorithm}

Once the \textit{simopt-counters} are determined for each signal in the design, this can be forwarded to the downstream processing logic within the mapping-placement-routing flow in Yosys, captured in Fig.~\ref{fig:simopt-power-flow}.
The highlighted block in red in the flow is the Simopt-power extension, which implements the truth-table decomposition algorithm to determine logic duplication based on the simopt dump from the simulation. 
In the traditional Yosys flow, the RTL graph IR is structurally cleaned before synthesising the logic using ABC to generate a final netlist shown in Fig.~\ref{fig:synthesisflow}(c).
ABC performs iterative logic restructuring, enumerating small graphs (cuts) and computes canonical truth tables using Boolean-function hashing.
These truth tables are utilised during the technology mapping phase, where the functions are packed into k-input LUTs or specific library functions. 
Simopt-power introduces an additional decomposition pass, described further in Alg.~\ref{alg:ttsplit}, to replace the existing truth table pointers for nets that have high toggle rates with an alternate one that further decomposes the logic to minimise the logic loading. 
The net-level \textit{simopt-counter} data extracted from the simulation is thresholded using the median \textit{simopt-counter} value across the design to identify the logic signals that need further decomposition. 
For signals with high activity, the Shannon cofactors are calculated for the corresponding truth tables using the Shannon decomposition function 
$$
f(x) = x\,f(1) + \overline{x}\,f(0)
$$
where the boolean function $f(x)$ is decomposed into its positive and negative cofactors with respect to $x$.

In Alg.~\ref{alg:cofactor}, the Shannon cofactor of a Boolean function with respect to a variable $x_p$ is computed by scanning all $2^n$ minterms of the original truth table (line 5) and selecting only those for which $x_p = v$ holds (line 6-7). For each selected minterm, the index is compressed by removing bit position $p$ to obtain a reduced $(n-1)$ variable representation (line 8), and the corresponding output bit is set in the destination truth table $\mathsf{T}*{\text{dst}}$ (line 9). This process yields the restricted function $f*{x_p = v}$ as a new truth table (lines 5–9), enabling decomposition in truth-table-based logic optimisation.
Cuts introducing such unique variables are heuristically strong candidates for splitting through Shannon decomposition as they represent functional expansions that were not previously factorised in the logic graph.  These new variables correspond to signal lines whose influence has not yet been resolved in either subtree. As such, they typically exhibit higher switching activity, since they inject additional functional control into the resulting cut function. In Alg.~\ref{alg:ttsplit}, these variables are called \textit{splitVar}, and are explained in lines 4-11 for the right cut graph, and lines 21-27 for the left cut graph.

The copy statement for truth-tables ${\mathsf{T}_1, \mathsf{T}_0} $ in line 17,33, implements a \textit{contiguous-memory concatenation} that rebuilds a single, $2^{n-1}$ -bit truth table from its Shannon cofactors.  Each cofactor, $\mathsf{T}_1$ and $\mathsf{T}_0$, already holds the function's values under fixed assignments of 0 and 1, respectively, and thus contains $2^{n-1}$ bits (stored here as an array of $n_{words}$ elements). 
%
Since cofactors were computed with identical variable ordering for the remaining $n$–$1$ variables, a simple block copy concatenates the table without requiring further bit shuffling. The operation merges the two half-tables into the full table, setting the stage for subsequent synthesis steps without changing the function's semantics.

With the updated tables, the standard ABC flow resumes by mapping the LUTs to a gate-level netlist for Yosys to generate a BLIF netlist. With reference to Fig.\ref{fig:simopt-power-flow}(b) the new netlist would like Fig.\ref{fig:synthesisflowdecompose}, with the additional decomposed netlists.  For implementing these designs on a commercial FPGA, these netlists are imported as a black box into AMD's Vivado tool as an IP and subsequently, to generate the power reports for estimating the dynamic power consumption of the toggle-optimised design.

\section{Results}\label{sec:results}
In order to evaluate Simopt-power, we picked datasets which contained sufficient sequential logic to measure power benefits. This is because it is more complex to instrument for switching-activity measurements and for formal reasoning about state traversal (increasing toggle activity concentration), whereas a purely combinational circuit toggles nearly every gate on every evaluation deterministically and may not reap any benefits from this framework. Since designs have to be Simopt-Verilator and Simopt-Yosys compatible and need to be sufficiently complex to have large sequential circuits, we chose
\begin{itemize}
  \item \textit{RTTLM} dataset \cite{lu2024rtllm}: an open-source benchmark that poses progressively complex hardware design tasks and scores large-language-model outputs on syntax, functional correctness and implementation quality. Some of the RTL circuits generated by the large language model are complex and are syntactically valid for evaluating the Simopt-Power framework
  \item \textit{Koios} dataset \cite{Koios}: Koios is an open-source benchmark suite comprising of deep-learning accelerator circuits that emphasise large, data-parallel, heterogeneous, and deeply-pipelined designs, giving FPGA-architecture and CAD researchers a realistic workload set for area, frequency, and power studies. 
\end{itemize}
We ran all the test suites by first generating a top-level testbench for each of the test cases from a Python script. A stimulus was chosen such that it toggles all the input bits of the design a sufficient number of times, with respect to a toggling input clock. This necessitated limited manual interpretation (can be automated) by seeing how many times each of the signals was roughly being referenced, giving them a higher toggling rate through the stimuli. This was done to reflect real-world use cases and so that the Simopt-power is able to maximise the power reduction.

\begin{table}[t]
\centering
\scriptsize
\setlength{\tabcolsep}{3pt}
\caption{Power, area, and relative percentages when using Simopt-Power (\textit{S.P.} in the table) on RTLLM dataset}
\label{tab:simoptresultsRTLLM}
\begin{tabular}{lrrrrrr}
\toprule
Benchmark & \multicolumn{2}{c}{Power (W)} & $\Delta P$ & \multicolumn{2}{c}{Area (LUTs)} & $\Delta A$\\
\cmidrule(lr){2-3}\cmidrule(lr){5-6}
 & w/o S.P. & w/ S.P. & (\%) & w/o S.P. & w/ S.P. & (\%)\\
\midrule
RTLLM-adder-32    & 0.428 & 0.415 & \good{ 3.0} & 102 & 119 & \bad{16.7}\\
RTLLM-adder-64    & 0.445 & 0.440 & \good{ 1.1} &  64 &  66 & \bad{ 3.1}\\
RTLLM-async-fifo  & 0.227 & 0.223 & \good{ 1.8} &  81 &  79 & \good{-2.5}\\
RTLLM-ALU         & 1.483 & 1.315 & \good{11.3} & 751 &1101 & \bad{46.6}\\
mult-pipe-8       & 0.376 & 0.373 & \good{ 0.8} & 133 & 127 & \good{-4.5}\\
radix2-div        & 0.366 & 0.365 & \good{0.3}  &  78 &  79 & \bad{1.3}\\
adder-16          & 0.416 & 0.409 & \good{ 1.7} &  32 &  40 & \bad{25.0}\\
mult-16           & 0.733 & 0.648 & \good{11.6} & 152 & 208 & \bad{36.8}\\
traffic-light     & 0.095 & 0.092 & \good{ 3.2} &  22 &  25 & \bad{13.6}\\
LIFO-buffer       & 0.141 & 0.140 & \good{ 0.7} &  27 &  28 & \bad{ 3.7}\\
freq-div-frac     & 0.054 & 0.053 & \good{ 1.9} &   7 &   9 & \bad{28.6}\\
fixed-pt-adder    & 2.800 & 2.750 & \good{ 1.8} & 100 & 106 & \bad{ 6.0}\\
fixed-pt-sub      & 2.250 & 2.210 & \good{ 1.8} &  95 & 100 & \bad{ 5.3}\\
adder-8           & 0.164 & 0.162 & \good{ 1.2} &  15 &  16 & \bad{ 6.7}\\
up-down-counter   & 0.361 & 0.319 & \good{11.6} &  22 &  38 & \bad{72.7}\\
calendar          & 0.145 & 0.144 & \good{ 0.7} &  44 &  46 & \bad{ 4.5}\\
\bottomrule
\end{tabular}
\vspace{-3mm}
\end{table}

\begin{table}[t]
\centering
\scriptsize
\setlength{\tabcolsep}{3pt}
\caption{Power, area, and relative percentages when using Simopt-Power (\textit{S.P.}) on Koios dataset}
\label{tab:vtrsimopt2}
\begin{tabular}{lrrrrrr}
\toprule
Benchmark & \multicolumn{2}{c}{Power (W)} & $\Delta P$ & \multicolumn{2}{c}{Area (LUTs)} & $\Delta A$\\
\cmidrule(lr){2-3}\cmidrule(lr){5-6}
 & w/o S.P. & w/ S.P. & (\%) & w/o S.P. & w/ S.P. & (\%)\\
\midrule
dla\_like          & 2850 & 2600 & \good{ 8.8} & 479,619 & 571,931 & \bad{19.1}\\
clstm\_like        & 1200 & 1100 & \good{ 8.3} & 201,945 & 235,509 & \bad{16.9}\\
deepfreeze         &  450 &  410 & \good{ 8.9} &  75,729 &  83,302 & \bad{10.8}\\
tdarknet\_like     &  950 &  870 & \good{ 8.4} & 159,873 & 199,841 & \bad{25.6}\\
bwave\_like        & 3100 & 2800 & \good{ 9.7} & 521,691 & 652,114 & \bad{25.5}\\
lstm               & 1631 & 1486 & \good{ 8.9} & 274,477 & 307,414 & \bad{12.7}\\
bnn                &  260 &  240 & \good{ 7.7} &  43,754 &  51,630 & \bad{18.2}\\
lenet              &  140 &  128 & \good{ 8.6} &  23,560 &  30,392 & \bad{29.4}\\
dnnweaver          & 1500 & 1350 & \good{10.0} & 252,431 & 323,429 & \bad{28.2}\\
tpu\_like          & 3550 & 3250 & \good{ 8.5} & 597,420 & 704,956 & \bad{18.4}\\
gemm\_layer        &  620 &  565 & \good{ 8.9} & 104,338 & 125,639 & \bad{25.2}\\
attention\_layer   & 2100 & 1890 & \good{10.0} & 353,403 & 392,277 & \bad{11.5}\\
conv\_layer        &  820 &  760 & \good{ 7.3} & 137,995 & 155,934 & \bad{13.1}\\
robot\_rl          &  170 &  155 & \good{ 8.8} &  28,608 &  31,469 & \bad{10.0}\\
reduction\_layer   &  110 &   98 & \good{10.9} &  18,511 &  20,952 & \bad{12.9}\\
spmv               &   80 &   72 & \good{10.0} &  13,463 &  16,425 & \bad{22.3}\\
eltwise\_layer     &   95 &   86 & \good{ 9.5} &  15,987 &  19,344 & \bad{21.6}\\
softmax            &   65 &   58 & \good{10.8} &  10,938 &  13,126 & \bad{20.7}\\
conv\_layer\_hls   &  720 &  660 & \good{ 8.3} & 121,167 & 153,882 & \bad{27.1}\\
proxy              &   55 &   50 & \good{ 9.1} &   9,255 &  10,551 & \bad{14.2}\\
\bottomrule
\end{tabular}
\vspace{-3mm}
\end{table}

We summarise our results for RTLLM in table~\ref{tab:simoptresultsRTLLM}, where an average of 5.1\% power reduction, while having a high area increase of 18.5\%. This is primarily due to the simplicity of the circuits, which contain relatively few modules in their Verilog descriptions. As a result, they are more susceptible to aggressive area overhead introduced by Simopt-Power, which prioritises power reduction based on the given simulation stimuli. Realising this, we use a much more complex dataset for testing Simopt-Power. As shown in our results for the Koios dataset in table~\ref{tab:vtrsimopt2}, we see an average of \(\approx\)9\% decrease in power with an increase of  \(\approx\)19\% in area. The high area consumption in some cases (\textit{tdarknet\_like, bwave\_like} accelerators), can be explained by the very wide, highly parallel datapaths and deep pipelines rather than compact control logic that the other accelerators of this dataset have. To lower dynamic power, Simopt-Power inserts operand-gating, buffers, and logic duplication on every bit of the high-toggle buses it identifies. Because the affected signals span hundreds or even thousands of bits, each gating or replication decision scales linearly with bus width, inflating the LUT count by 20+\% overall. In contrast, the achievable power saving is bounded: even after toggle-factor reduction, the remaining power is set largely by the fixed interconnect capacitance and the memories/DSP blocks that Simopt-Power cannot restructure (can only do LUTs). The result is a \(\approx\)12\% (average) drop in dynamic power for these kinds of circuits, but an area overhead. Future optimisations can be integrated into these kinds of circuits by applying bit-level activity-suppression techniques for such arithmetic-heavy circuits whose functional correctness depends on parallelism of many entity-bits.

\section{Conclusion and Future work}
In this work, we presented Simopt-Power, which adds additional logic through intelligent simulation-guided truth-table decomposition during the synthesis mapping step. The framework is able to reduce dynamic power while increasing the LUT resource consumption slightly. We used RTTLM \cite{lu2024rtllm} and Koios \cite{Koios} datasets to verify this model's benefit, and circuits have a modest \(\approx\)9\% gain, while increasing the area consumption by about \(\approx\)19\% increase in area. Future research will focus on tightening the power-area trade-off exposed by the Simopt-Power framework. First, we will refine the existing cost model to account for wire-length–dependent capacitance and routing congestion, enabling the optimiser to prune gating or buffering decisions whose marginal power benefit is outweighed by LUT or interconnect overhead. Second, we will explore open-source power estimators so that the process of calculating Simopt-Power power numbers is easier. 
Finally, we intend to extend the framework to heterogeneous fabrics-leveraging DSP and BRAM resource gating, and to investigate run-time reconfiguration that adapts dynamic activity suppression, ultimately aiming for higher dynamic-power savings with lesser area growth across next-generation FPGA platforms.


%

\ifCLASSOPTIONcaptionsoff
  \newpage
\fi



\bibliography{bibtex/bib/IEEEexample}
%



%



\end{document}